\documentstyle[aps,twocolumn,floats,epsf]{revtex}

\def\celcius{$^{\circ}$C}

\begin{document}

\draft

\twocolumn[\hsize\textwidth\columnwidth\hsize\csname@twocolumnfalse\endcsname
\title{Impurity mediated nucleation in hexadecane-in-water emulsions}
\author{Amy Herhold, Deniz Erta{\c s}, Alex J.\ Levine\cite{home}
and H. E. King, Jr.
}
\address{Exxon Research \& Engineering Co., Rte 22 E. Annandale NJ 08801}
\date{\today}
\maketitle
\begin{abstract}
We report detailed nucleation studies on the liquid-to-solid transition
of hexadecane using nearly monodisperse hexadecane-in-water emulsions.
A careful consideration of the kinetics of isothermal and nonisothermal
freezing show deviations from predictions of classical nucleation theory,
if one assumes that the emulsion droplet population is homogeneous.
Similar deviations have been observed previously\cite{Turnbull:61}. As an
explanation, we propose a novel argument based on the dynamic generation of
droplet heterogeneity mediated by mobile impurities. This proposal is in
excellent agreement with existing data.
\end{abstract}

\pacs{PACS numbers: 82.60.Nh, 64.60.Qb,44.60.+k,82.70.Kj}

]

\section{Introduction}
\label{sec:intro}
The use of micron-sized emulsified droplets for nucleation studies of the
liquid-to-crystal transition is a well established technique.  Vonnegut
\cite{Vonnegut:48} pioneered this concept in the first half of this century
and others have used it to  study a variety of
materials\cite{Turnbull:52,Turnbull:61,Kelton:91,Perepezko:97}. The
advantage of using  a large ensemble of independent nucleation sites to
measure the stochastic process of nucleation is obvious.  An additional
advantage is that the effects of crystal growth are virtually eliminated.
On the time scale of the nucleation measurements, the growth time for each
droplet is instantaneous.  Furthermore, because experiments are conducted
in a range where the probability of two nucleation events per drop is
small, it is a simple matter to equate the total number of crystallized
droplets with the number of nucleation events.

The role of impurities on nucleation in emulsion studies is not obvious.
Clearly, one possible role is to act as heterogeneous nucleation sites.
Both Turnbull\cite{Turnbull:52} and Perepezko\cite{Perepezko:84} have
discussed this.  A signature of heterogeneous nucleation is a small
difference between the melting temperature and the onset of nucleation.
For example, in mercury, Turnbull showed that changes in surfactants could
increase the undercooling from 5 \celcius\ to 60 \celcius, with the smaller
values attributed to the effects of heterogeneous nucleants either on the
surface or within the volume of the droplets. It is well known that
emulsification tends to increase the undercooling over that of bulk
liquids, but a simple calculation shows that for this to result from
isolation of the heterogeneous catalysts they must be present at extremely
low levels (cf. Sec. \ref{sec:nonuniform}). This seems to suggest that most
impurities are not effective nucleating agents\cite{Perepezko:97}.

A second role of impurities is to lower the melting point.
Perepezko\cite{Perepezko:84} has shown in his studies of metals that as the
liquidus temperature drops due to alloying, there is a corresponding
decrease in the nucleation temperature.  Hence an approximately constant
value of the undercooling is observed. These effects can be large, for
example in the lead-antimony system the liquidus and nucleation
temperatures both decline by about 75 \celcius\ as the antimony content
increases up to $17.7$ atomic percent.  The effect of a typical impurity
level (mole fraction $\approx 0.01$) will of course be much smaller.
However, the nucleation rate is strongly temperature dependent. For
example, in n-alkanes (C$_{n}$H$_{2n+2}$, henceforth abbreviated C$n$), the
nucleation rate changes by a factor of 5000 per \celcius\cite{Turnbull:61}.
We show in this work that impurity levels of even a few percent cause a
significant change in nucleation behavior. This influence must be accounted
for to extract an accurate value of the nucleation rate.

Size of the emulsion droplets also plays a key role in nucleation studies.
In homogeneous nucleation, the nucleation rate is proportional to the
volume of the droplets. Typically, the determination of the size
distribution for the emulsions is a large source of error in nucleation
rate measurements\cite{Turnbull:61}. Advances in emulsion synthesis
techniques now makes it possible to create nearly uniformly sized emulsion
particles.  The average deviation in diameter from the mean size is only
from $10$ to $15 \%$\cite{Bibette:91,Kandori91a,Kandori:91}. By using such
emulsions with narrow size distributions, the determination of the rate
constant for any given emulsion radius is considerably improved. Also, this
makes it possible to better test the predicted volume scaling of the
nucleation rate.

In this paper, we report the results of a refined experimental and
theoretical investigation of the nucleation rate in emulsified hexadecane
(C16). The quality of experimental data has been significantly improved by
using nearly monodisperse emulsions in a well controlled thermal
environment and by using x-ray scattering to accurately monitor the volume
of nucleated droplets during crystallization. Data were obtained for both
fixed and linearly increasing undercooling as a function of time,
henceforth referred to as isothermal nucleation and linear cooling,
respectively. Thermodynamic melting curves have also been obtained for the
samples in order to assess the influence of impurities introduced during
the emulsification procedure. The theoretical analysis has concurrently
been refined to account for the {\it entire} time-dependence of the solid
fraction as a function of time, rather than just matching some
characteristic time. This careful analysis revealed deviations from a
simple scenario involving a uniform ensemble of independent nucleation
events. By systematically relaxing the assumptions in this description, we
have concluded that the most probable cause of the observed deviations is a
novel mechanism which involves the transport of impurities expelled from
nucleated droplets to the remaining liquid droplets. This transport
subsequently increases the impurity concentration of these remaining
droplets and reduces their nucleation rate by lowering their thermodynamic
melting point $T_m$. This mechanism can completely account for the observed
behavior and can be further explored with additional experiments.

The remainder of this paper is organized as follows: In
Sec.~\ref{sec:review}, we review previous studies on alkane nucleation.
Sec.~\ref{sec:exp} provides details of the experimental techniques and
procedures used. Sec.~\ref{sec:uniform} reviews classical nucleation theory
and its predictions based on the assumption of uniform nucleation, i.e.,
independent nucleation events in a homogeneous ensemble of droplets.
Finding that this theory is unable to account for the observed behavior of
the system, we consider possible extensions of this theory in
Sec.~\ref{sec:nonuniform}. Here we explore two scenarios for the generation
of droplet heterogeneity, which is necessary to explain the data. These
include the possibility of some fixed-in-time heterogeneity generated by
the emulsification process, such as a distribution of droplet sizes or
impurity concentrations, and the dynamic generation heterogeneity in the
droplet population due to the nucleation process itself. Based on the
reanalysis of the data, we present our conclusions in Sec.~\ref{sec:concl}
and suggest a number of experiments to further test the validity of the
proposed mechanism, as well as avenues for future theoretical
consideration.

\section{Previous studies of alkane nucleation}
\label{sec:review}
Four groups have studied alkane nucleation through the use of emulsion
samples. The earliest work is from Turnbull and Cormia \cite{Turnbull:61}
who studied C16, C17, C18, C24 and C32. Theirs is the first evidence that
alkane nucleation is unusual. First, they noted that there seemed to be an
unusual spread in the melting temperatures.  This was characterized as
``sharp'' and ``broad'' melting fractions.  The amount of each varied from
sample to sample, even for the same chain length.  Because they performed
isothermal nucleation studies they were sensitive to this spread in melting
temperature.  To analyze the nucleation behavior in this situation they
focused on the early-time data with transformed fraction $n < 0.5$. This
narrowed focus was meant to isolate the behavior of the sharp-melting,
majority phase.

The second anomaly in alkane nucleation is the ease with which the alkanes
nucleate.  Stated in terms of reduced undercooling, $\Delta T_r=
(T_{m}-T_{N})/T_{m}$, where $T_N$ is the point where the nucleation rate
becomes significant and $T_{m}$  is the thermodynamic melting temperature,
$\Delta T_r$ for the alkanes is about 0.05 whereas that for other materials
is from 0.2 to 0.5 \cite{Turnbull:78}. Turnbull and Cormia's analysis of
the nucleation behavior in terms of the classical nucleation model showed
that the nucleation barrier is small ($9.64$ mJ/m$^{2}$ for C18; detailed
kinetics measurements were not performed for C16) but that the
pre-exponential factor is in rough agreement with that calculated from
classical nucleation theory (experimental value $=10^{37.35\pm 2}$
m$^{-3}$s$^{-1}$ for C18). The small barrier accounts for the small
undercooling temperature.  The agreement of the pre-exponential with that
from classical nucleation theory is itself unusual.  Other materials
typically exhibit values several orders of magnitude larger in value than
the classical value \cite{Kelton:91}.

Two subsequent studies on alkane nucleation explored the effect of changing
carbon number \cite{Uhlmann:75,Oliver:75}. In both studies continuous
cooling data were used to calculate the crystal-liquid surface tension and
the pre-exponential factor.  Both studies found that as the carbon number
is reduced below about $n_C=15$, there is an increase in the barrier.  The
pre-exponential values were not precisely determined by these studies.
Re-analysis in terms of the Negentropic Model \cite{Turnbull:78} led to a
value of $9$~mJ/m$^{2}$ for the barrier of C16, and to the suggestion that
the anomalous barrier height disappears as $n_C \longrightarrow  1$ and as
$n_C \longrightarrow \infty$, and that its origin is in partial alignment
of the alkane chains.

The fourth group
\cite{McClements:90,Dickinson:93,Dickinson:90,Dickinson:91} to study
nucleation in emulsions focused on the behavior of C16. They utilized
ultrasound transmission to measure the proportion of liquid to solid in an
emulsion sample.  Both stepwise cooling and isothermal hold experiments
were used.  Their results exhibit the typical  $\sim
14-15^{\mbox{\tiny{o}}} {\rm C}$ undercooling found by other workers,
showing that their samples behave as expected during stepwise cooling.
However, the real focus of these studies was the interaction of liquid and
crystalline droplets during isothermal hold experiments
\cite{McClements:90}.  At an undercooling of $\sim 10$ \celcius, where the
nucleation rate for the C16 liquid droplets alone is quite small, equal
volume mixtures of liquid and solid C16 droplets were studied. The solid
C16 drops were created by an initial deep undercooling and then physically
mixed together with the as-yet unfrozen sample.  The authors found that the
solid C16 particles accelerate nucleation. The particle size distribution
before and after the experiment was similar; therefore they ruled out
Ostwald ripening as the source of the acceleration.  No attempt was made to
fit this data to a rate law.

In a set of related experiments the type and concentration of the
surfactant was varied \cite{Dickinson:93}. The presence of crystalline C16
droplets accelerated the nucleation by varying amounts depending on the
surfactant type. It was also found that the rate of nucleation for the
mixed solid/liquid emulsion increased as the amount of surfactant
increased.  The authors suggest that inter-particle collisions are
responsible for the accelerated nucleation.

As we discuss in Sec.~\ref{sec:uniform}, we observe an effect opposite to
that seen by these authors.  As our sample crystallizes, the remaining
crystallization is more difficult. However, Dickinson {\it et al.\/}
\cite{Dickinson:91}  also report an experiment where this effect is seen.
In this work they hold their emulsion at a undercooling of about 15
\celcius\ and follow the percent transformed with time.  Like us, they
observe an initially rapid nucleation which then slows.  They do not
attempt to fit this to a rate law, but it is clear that a single
exponential cannot describe their data.  Also similar to our findings they
report that the melting behavior extends over a large temperature regime.
They report melting ranges of about 2 \celcius\ for all their samples, and
in one instance \cite{Dickinson:91} they report the onset of melting more
than 10~\celcius\ below the bulk melting temperature.  As we show in
Sec.~\ref{sec:impurity}, this is a signature of the impurity effects which
can affect nucleation.

\section{Experimental Procedure}
\label{sec:exp}

\begin{figure}[tbp]
\epsfxsize=0.95\columnwidth
\centerline{\epsfbox{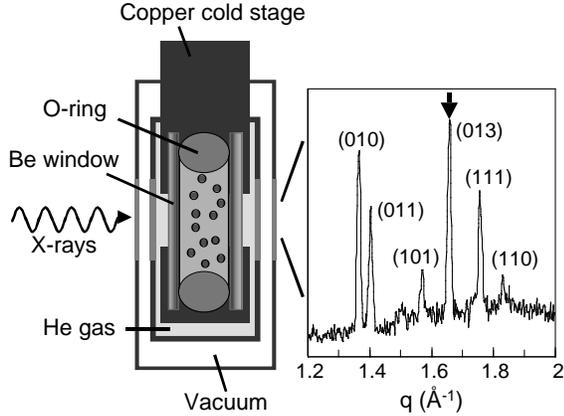}}
\caption{{\it Left:} A schematic of the x-ray cell
loaded with an emulsion sample. {\it Right:} The powder pattern of the C16
emulsion sample in the triclinic (solid) phase. Numbers indicate the Miller
indices
for each peak. The arrow denotes the detector position, which
monitors the (013) peak to determine the amount of solid in the sample
during the kinetics measurements.}
\label{setup}
\end{figure}

Nucleation of the liquid to crystal transition was measured in droplets of
C16 of an oil-in-water emulsion. The observed crystal phase is the
thermodynamically stable triclinic phase. A schematic of the experimental
setup is shown in Fig.~\ref{setup}. A reference table of relevant material
properties for C16 is given in Table~\ref{c16table}.

\begin{table}
\begin{tabular}{l|r}
Melting Temperature: $T_{m0}$\quad & $291.32$ K \\ \hline
Entropy of Fusion: $\Delta S$ & $6.28 \times 10^5 \mbox{J} \, \mbox{m}
^{-3} \mbox{K}^{-1}$ \\ \hline
Mass Density: $\rho$ & $773.4 \mbox{kg} \, \mbox{m}^{-3}$ \\ \hline
Molar Weight: $M_w$ & $0.22643 \mbox{kg} \, \mbox{mol}^{-1}$ \\ \hline
Viscosity at $T_{m0}$: $\eta $ & $3.484 \times 10^{-3} \mbox{Pa s}$
\end{tabular}
\caption{Material Properties of Hexadecane\protect \cite{tableref}}
\label{c16table}
\end{table}

One $\mu m$ diameter emulsions of C16 in water with the surfactant sodium
dodecyl sulfate (SDS) were prepared using the fractionation method of
Bibette\cite{Bibette:1991}. The distribution of droplet sizes was less than
$\pm 10 \% $ in diameter as determined by light scattering and optical
microscopy.

Below its Krafft point at 283 K\cite{Krafftpoint}, SDS precipitates out of
solution as a crystalline solid in equilibrium with a small concentration
of dissolved monomers.  In order to avoid this effect, a co-surfactant was
added to supress the Krafft point\cite{EOref}.  The chosen surfactant, a
sulfated polyoxyethylenated alkyl alcohol (refered to here as EO), has the
chemical formula
CH$_{3}$(CH$_{2}$)$_{11-15}$(OCH$_{2}$CH$_{2}$)$_{9}$OSO$_{3}$Na.  An
aqueous solution of EO was added to the initial emulsion sample to achieve
approximately equal concentrations of SDS and EO in the range of 0.5 - 1
wt$\%$. This new solution was allowed to equilibrate for more than 24 hours
before use, during which time the emulsion droplets had flocculated
together to form a white cream floating on the aqueous phase. The cream of
the emulsion sample with both SDS and EO co-surfactants was loaded directly
into the x-ray cell.

\begin{figure}[tbp]
\epsfxsize=0.95\columnwidth
\centerline{\epsfbox{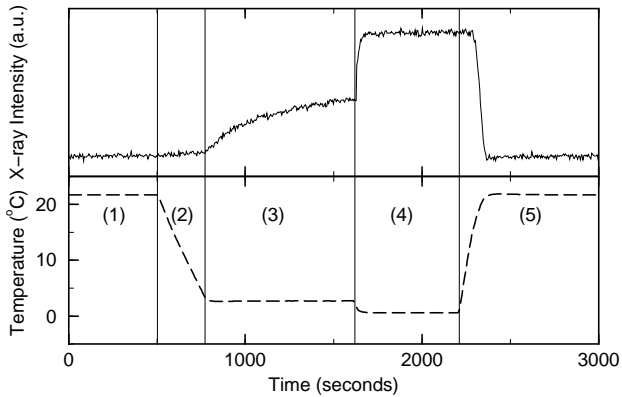}}
\caption{The temperature of the sample and measured x-ray intensity as a
function of
time over the course of a typical isothermal experiment. The sample was
first held
above $T_m$ to ensure that all droplets were liquid (1). The sample was then
quenched to (2) and held at (3) a fixed temperature $T$, and the rise in
scattering
intensity was measured. A subsequent quench to a lower temperature (4) was used
to determine the scattering from a fully crystallized sample. The sample
was then heated back above $T_m$ (5).}
\label{experimental}
\end{figure}

The emulsion sample was loaded into a Viton O-ring sandwiched between two
Be windows, held between two copper plates (see Fig~\ref{setup}). The
sample temperature was read with thermistors ($\pm 0.2$ K accuracy) placed
in the copper less than 3.5 mm from the sample. The sample was cooled on an
Air Products Displex cold finger with a closed-cycle He refrigeration
system. Heating wire wrapped around the base of the cold finger provided
temperature control. As the cold finger was operated in vacuum, an
air-tight copper shield with kapton windows was placed directly around the
copper sample mount inside the vacuum shroud of the Displex to surround the
sample with gas and therefore reduce temperature gradients.  The
temperature was controlled with a LakeShore 340 Temperature Controller. The
control temperature sensor (Si diode) was embedded in the cold finger
directly above the copper sample mount.  Using this setup, it was possible
to cool the sample from room temperature down to 277 K in less than 250 s
without any overshoot of the temperature (Fig.~\ref{experimental}). This
careful attention to temperature gradients overcame problems encountered in
the initial studies \cite{HenryShao}.

The nucleation rate was determined by measuring the change in intensity of
the x-ray diffraction as a function of time. X-ray scattering was
performed using a Rigaku 18~kW rotating anode generator and an x-ray
wavelength of 1.542 \AA. Scattered x-rays were detected with a Bicron
detector on a Huber 4-circle
spectrometer. Scattering from the solidified 1 $\mu m$ dia. droplets
resulted in a powder diffraction pattern of the triclinic phase.
The (013) Bragg peak of the triclinic phase was selected to be monitored
(Fig~\ref{setup}) due to the intensity of the peak and the low scattering
from the liquid
phase at that angle. For the isothermal
experiments, the detector slits were set very wide to increase the signal
intensity and to encompass slight
differences in the peak position due to thermal expansion when measuring at
different temperatures. During the constant cooling rate and the melting
experiments, the slits were narrowed and the detector was continually
shifted to track the
peak during thermal expansion.

Three types of experiments were conducted: crystallization at a constant
undercooling (isothermal nucleation), crystallization while the temperature
was lowered linearly in time (linear cooling), and melting at a slow
heating rate.  For the isothermal experiments, the sample was rapidly
quenched from room temperature down to the desired undercooling and held at
that temperature while the increase in x-ray scattering was followed as a
function of time (see Fig.~\ref{experimental}). In order to determine the
total scattering signal when the entire droplet population was frozen, each
experiment was followed by a deep quench to a lower temperature (273.8 K)
that froze all droplets. All reported isothermal data is normalized to the
final intensity value of the subsequent deep quench.  In addition, all
quenches were repeated multiple times with identical results.

In the linear cooling experiments, the rate of temperature change was
between $-0.01$ and $-0.002$ K/min. Faster rates were avoided because the
latent heat released by the crystallizing droplets could not be dissipated
quickly enough and resulted in a smearing of the transformation over a
larger temperature range.

Melting was done in steps of 0.25 to 1 K.  The sample was heated slowly
between steps and was allowed to equilibrate for approximately 1 hour at
each step. This equilibration was observed by the leveling out of the
intensity subsequent to each temperature step.

\section{Uniform Nucleation}
\label{sec:uniform}

In this section, we interpret the results of the experiments using classical
nucleation theory, with the assumptions that (i) all droplets in the
emulsions are
identical, and that (ii) nucleation events are statistically independent in
each droplet. We refer to this set of assumptions as uniform nucleation.

In classical nucleation theory \cite{Kelton:91,Turnbull:49}, the formation
of the thermodynamically stable
crystalline phase in an undercooled, metastable liquid phase is controlled by
the local free energy barrier towards solidification. The free energy
cost, $\Delta G(R)$, associated with the creation of a solid sphere of
radius $R$ is given by
\begin{equation}
\Delta G=-\frac{4\pi }{3}R^{3}\Delta g+4\pi R^{2}\gamma,
\label{free-energy}
\end{equation}
where $\Delta g$ is the free energy difference per volume between the
metastable
liquid and the stable solid phase, and $\gamma $ is the surface tension
between the
two phases. For small solid radii, $R$, the second term dominates the first and
the formation of the more stable solid phase requires an increase in free
energy.
The nucleation rate is controlled by the free energy barrier associated
with the formation of a solid sphere at a critical radius of
$R_{c}=2\gamma /\Delta g$, since further solidification reduces the free energy
of the system. For small undercooling, the thermodynamic driving force
$\Delta g=\Delta S\,\Delta T$, where $\Delta S$ is the entropy of fusion
of the phase transition and $\Delta T=T_{m}-T$ is the degree of
undercooling of the
sample at temperature $T$\cite{cp-note}. We neglect the dynamics of the
growth and simply assume that upon formation of a solid phase of critical
radius, the entire droplet freezes instantaneously.

The probability of spontaneously creating a solid with the critical radius
is simply found from Eq.(\ref{free-energy}), leading to a nucleation rate
\begin{equation}
\nu _{o}=AV\exp \left( -\frac{\Delta G(R=R_{c})}{k_{\mbox{\tiny{B}}}T}%
\right) .  \label{rate}
\end{equation}
This rate is proportional to the volume, $V$, of available undercooled fluid,
the Boltzmann factor associated with the free energy barrier for creating a
critical nucleus as determined from Eq.(\ref{free-energy}), and an attempt
frequency per unit volume, $A$, for creating such a critical nucleus, which
is given by
\cite{Wu:96}
\begin{equation}
A=C\omega ^{1/3}\sqrt{\frac{\gamma }{k_{\mbox{\tiny{B}}}T}}\alpha \frac{D}{
d^{2}}.  \label{prefactor}
\end{equation}
Here, $C$, $\omega $, $\alpha $, $d$ and $D=(k_{\mbox{\tiny{B}}}T)/(3\pi
\eta d)$ are, respectively, a numeric constant $C=1.65$, the monomeric
volume, molecular number density, molecular diameter and the
Stokes-Einstein diffusion constant of C16 in its liquid phase. For
notational simplicity we define
\begin{equation}
\Omega \equiv \frac{16\pi \gamma ^{3}}{3k_{\mbox{\tiny{B}}}T_{m}\Delta S^{2}},
\label{exp}
\end{equation}
so that the nucleation rate given by Eq.(\ref{rate}) becomes
\begin{equation}
\nu _{o}=AV\exp \left( -\frac{\Omega }{(\Delta T)^{2}}\right) .
\label{rate2}
\end{equation}

\subsection{Isothermal Nucleation}

At a fixed level of undercooling, the rate at which emulsion droplets freeze
is time independent. Let $n(t)$ represent the fraction of solid droplets
at time $t,$ following a rapid quench to a given undercooling $\Delta T$ at
$t=0$. Then $n(t)$ obeys the differential equation
\begin{equation}
\frac{dn}{dt}= \left(1 - n \right)\nu _{o},  \label{exponential}
\end{equation}
with initial condition: $ n(0) = 0$.
This differential equation may be integrated to give
\begin{equation}
n(t)= 1 - e^{-\nu _{o}t},
\label{exponential-ans}
\end{equation}
so the fraction of liquid droplets decreases exponentially in time.

\begin{figure}[tbp]
\epsfxsize=0.95\columnwidth
\centerline{\epsfbox{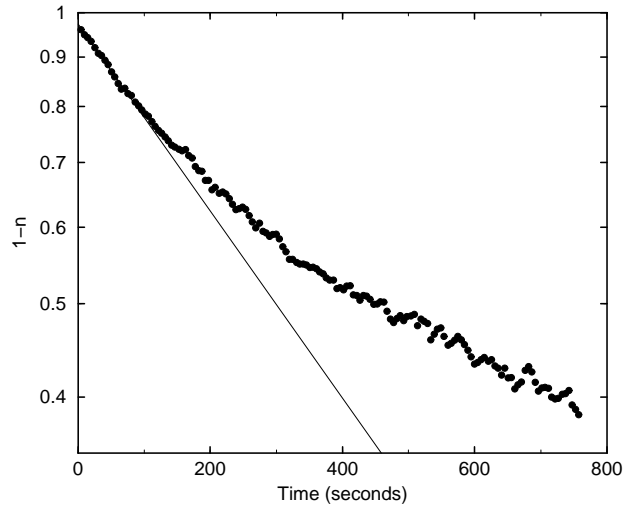}}
\caption{The fraction of liquid emulsion droplets as a function of time at
constant undercooling of 15.3 K as measured by the x-ray intensity at a
Bragg peak
of the crystalline C16 phase. If the nucleation rate had been
independent of time, this process should have yielded a simple
exponential dependence on time. The solid line represents the expected
behavior based on the nucleation rate at early times ($t<100$ s.)
The data deviates significantly from this simple model at later times.}
\label{nonexponential-pic}
\end{figure}

It should be noted that the rate $\nu _{o}$ depends strongly on both the
surface tension $\gamma $ of the liquid--solid interface and the degree of
undercooling $\Delta T$ [cf. Eqs.(\ref{exp})-(\ref{rate2})]. This strong
dependence allows an extremely precise determination of $\gamma $ provided
that the data can be fit to a {\em single} exponential. On the other hand,
Fig.~\ref{nonexponential-pic}, showing a representative data set from an
isothermal freezing experiment, demonstrates that such a simplistic
interpretation of the data is inapplicable. The freezing of remaining liquid
droplets at later times proceeds far slower than suggested by an
extrapolation from the initial decrease of the fraction of liquid droplets,
$\left( 1 - n \right) $,
shown by the solid line in Fig.~\ref{nonexponential-pic}.

\subsection{Linear Cooling}
\label{sec:lincool}

\begin{figure}[tbp]
\epsfxsize=0.95\columnwidth
\centerline{\epsfbox{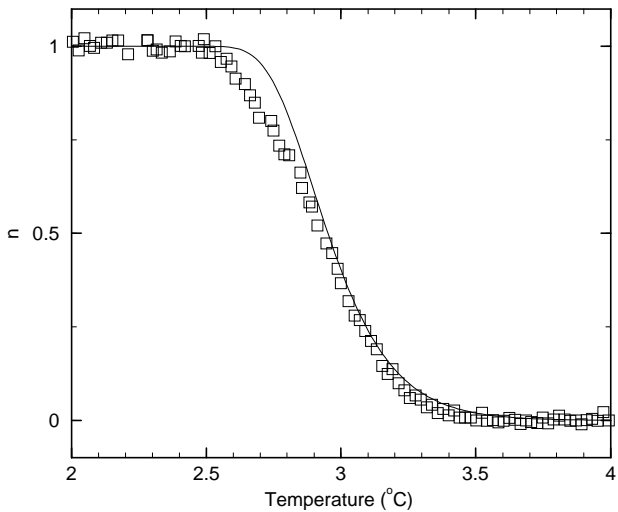}}
\caption{Solid fraction as a function of temperature at a constant
cooling rate, $\lambda= 4.9 \times 10^{-3}$ \celcius/min, plotted together
with the theoretical curve given by Eq.~(\protect{\ref{linear-ans}}).
Note that time evolves from right to left.}
\label{linear-pic}
\end{figure}

We now consider the case where the temperature, rather than being held
constant, decreases linearly in time. The thermodynamic driving force thus
increases in time and the exponential decay of the liquid fraction seen in
the solution of Eq. (\ref{exponential}) is changed to a relatively sudden
drop at an undercooling determined by the cooling rate (see Fig.~\ref
{linear-pic}).

Suppose that the temperature of the sample is decreased at a rate given by $%
\lambda $. If we ignore the weak temperature dependence in the
pre-exponential factor in Eq.~(\ref{rate2}), we find that the {\it %
time-dependent} nucleation rate is now given by
\begin{equation}
\nu (t)=AV\exp \left( -\frac{\Omega }{\left( T_{m}-T(0)-\lambda t\right) ^{2}}
\right).
\label{rate(t)}
\end{equation}
The differential equation describing the time evolution of $n(t)$ reads
\begin{equation}
\frac{dn}{dt}= \left(1 -n \right) AVe^{-\frac{\Omega }{\lambda ^{2}t^{2}}},
\label{linear_diff}
\end{equation}
where the initial time, $t=0$, is chosen to occur when the sample is at
$T_{m}$.

Following a change of independent variable to $z\equiv \Omega /\left(
\lambda t\right) ^{2},$ the solution of Eq.~(\ref{linear_diff}) is given by
\begin{equation}
n(t)=1 - \exp \left[ -\frac{AV\sqrt{\Omega }}{2 \lambda }
\Gamma \left( -\frac{1}{2},%
\frac{\Omega }{(\lambda t)^{2}}\right) \right] ,
\label{linear-ans}
\end{equation}
where we have introduced the incomplete Gamma function\cite
{Abramowitz:72}
\begin{equation}
\Gamma \left( -\frac{1}{2},z\right) \equiv \int_{z}^{\infty }
e^{-\tilde z} {\tilde z}^{-3/2}d{\tilde z}.
\label{int-rep}
\end{equation}
For $z \gg 1$, this function goes to zero as $%
z^{-3/2}\exp \left( -z\right) $ so at short times $n(t)$ behaves as
\begin{equation}
n(t)\sim \frac{AV\lambda ^2}{2\Omega }t^{3}e^{-\frac{\Omega }
{\lambda ^{2} t^{2}}}, \; t \ll \lambda/\sqrt{\Omega} .
\label{linear-ans-approx1}
\end{equation}
Thus, at $t=0$ the function has an essential singularity and the initial
increase in the solid fraction $n(t)$ is extremely slow. For $z \ll 1$,
the integral in Eq.(\ref{int-rep}) is dominated by the divergence of the
integrand at $z=0$ and is approximately given by $2 z^{-1/2}$, so at long
times
\begin{equation}
n(t) \sim 1 - \exp (-AVt), \; t \gg \lambda/\sqrt{\Omega },
\end{equation}
as expected since the free energy barrier disappears. However, since the
attempt frequency $AV$ is typically very large, the nucleation process is
largely finished long before the free energy barrier becomes negligible.
Therefore, this asymptotic behavior is not observable experimentally.
The complete solution is plotted in Fig.~\ref{linear-pic} together with the
experimental data.

The plot shown in Fig.~\ref{linear-pic} of Eq.~(\ref{linear-ans}) was made
using the pre-exponential factor taken from a fit to the isothermal data
(see Sec.~\ref{sec:return}) and using the surface tension $\gamma $ as a
fitting parameter. The pre-exponential factor $A$ used in this fit was
calculated at $T_{m}$ ($A$ at the temperature of transformation is less
than 3\% smaller than $A$ at $T_{m}$). We find that the best fit value of
this surface tension is $\gamma =9.9\pm 0.2$ mJ/m$^{2}$. It should be noted
that, to a good approximation, shifting the value of the surface tension
$\gamma $ simply translates the theoretical curve in time ({\it i.e.\/}
temperature). Therefore, $\gamma $ can in principle be determined with
great precision, since the time at which abrupt change in $n(t)$ occurs is
exponentially sensitive to its value. The principal source of the
uncertainty is the accuracy with which the magnitude of the undercooling
was measured. This result for the surface tension is in agreement with the
previous work of Turnbull {\it et al.} for C18 \cite{Turnbull:61} and other
groups for C16 \cite{Uhlmann:75,Oliver:75}.

Finally, we note that the agreement between the linear cooling data and the
prediction of Eq.~(\ref{linear-ans}) is not exact.  In particular the slope
of the data in the transition region (3.2 -- 2.6 \celcius\ in
Fig.~\ref{linear-pic}) is clearly smaller than that of the theoretical
prediction. This behavior is consistent with the observed slowdown of the
nucleation rate in isothermal experiments. In order to address these
discrepancies, we proceed in the next section to systematically relax the
assumptions made in the uniform nucleation hypothesis.

\section{Nonuniform Nucleation}
\label{sec:nonuniform}

Applying the results of uniform nucleation theory to both the isothermal
and linear cooling experiments has led to substantial disagreements between
theory
and experiment in the former and more subtle deviations in the latter.
The departure from simple exponential behavior in the isothermal quench
experiments
has been noted by previous workers in the field\cite{Turnbull:61,Dickinson:91}.
Faced with this observation, one is forced to consider one of two general
possibilities:

(i) The emulsion droplets are not homogeneous, i.e., there is a
``quenched-in" dispersion in some relevant property, such as a distribution of
sizes or impurity concentrations, leading to {\it multi-}\/exponential
decay of $n(t)$ in the isothermal experiments, or,

(ii) Nucleation events are not statistically independent, i.e., the nucleation
rate of a given emulsion droplet depends on the state of the other
droplets, such
as the current fraction of solid droplets in the emulsion, possibly due to an
impurity mediated interaction between the droplets. This sort of
interaction might
lead to a history dependent nucleation rate and result in non-exponential
freezing
curves.
We re-emphasize that the non-exponential
character of the curve (Fig.~\ref{nonexponential-pic}) does not reflect the
finite growth rate of the solid as this rate is extremely fast compared to the
data acquisition rate\cite{Turnbull:69,growth-note}. Similarily we may not
attribute
the deviation from a simple exponential to latent heat effects (rise in
temperature due to rapid release of heat from crystallization) since the
magnitude
of the deviation does not diminish with smaller uncoolings and hence, lower
nucleation rates. In addition, given that the isothermal curves can be
reproduced repeatedly for the same loading of a given sample, the
non-exponential character does not arise from coarsening of the size
distribution of the emulsion sample. The main difference between mechanisms
(i) and (ii) is that in
the former scenario, although there is a distribution of nucleation rates,
it does not depend on the state of the droplets, whereas the latter scenario
suggests that nucleation events influence the subsequent evolution of the
system.

Let us start by considering the first possibility, the existence of some fixed
in time distribution of nucleation rates. If there is a population of
droplets with different volumes, Eq.~(\ref{exponential-ans}) can easily be
generalized to account for this effect by integrating contributions to
the freezing curve of droplets of a given volume $V$ over the volume
distribution
 ${\cal P}(V)$. The resulting isothermal freezing curve is then
[cf. Eqs.(\ref{rate2}), (\ref{exponential-ans})]
\begin{equation}
\bar{n}(t)=1-\int dV {\cal P}(V)\exp\left(-AV e^{-\Omega / (\Delta T)^2}
\right).
\label{polydispersity}
\end{equation}
We take the droplet volume distribution to be log normal with characteristic
volume $V_{0}$ and variance $\sigma ^{2}:$\cite{Mason:98}
\begin{equation}
{\cal P}(V)=\frac{e^{-\sigma ^{2}/2}}{V_{o}\sqrt{2\pi \sigma ^{2}}}\exp
\left\{ -\frac{\left[ \log \left( \frac{V}{V_{o}}\right) \right] ^{2}}{%
2\sigma ^{2}}\right\} ,  \label{lognormal}
\end{equation}
and find that the best fit to the experimental data is obtained for $\sigma
\approx 1.0$. The resulting best fit to the data is poor; also, the emulsion is
actually known to have a much narrower volume distribution, based on the
method of emulsion preparation and our own optical measurements
\cite{emulsionsize-note}. Furthermore,
isothermal experiments at different undercoolings cannot be accounted for by a
single set of fitting parameters. Thus, we do not believe that the observed
non-exponential saturation of the isothermal freezing curves can be attributed
to the polydispersity of the emulsion sample.

Along a similar vein, one may postulate that there exists some fixed
distribution of melting temperatures in the emulsion droplet population,
possibly due to a distribution of impurity concentrations in the emulsion
droplets. Here one must assume that the impurities in the droplets are
highly insoluble in water, in order to prevent the equalization of impurity
concentration via diffusion through the continuous phase. Problems very
similar to the case of size distribution arise here as well. Although good
fits to individual isothermal nucleation curves can be obtained using this
modification, experimental runs at different undercoolings cannot be
accounted for using the same fixed impurity concentration distribution.
Furthermore, in order to fit a given data set one is once again forced to
assume a distribution of impurity concentrations that is too broad (with
standard deviation comparable to the mean). The average number of
impurities in a droplet of volume $V$ is $N_i=c \rho V N_A/M_w$, where $c$
is the mole fraction of impurities, $N_A$ is Avogadro's number, and $\rho
$, $M_w$ are the density and molar weight of C16, respectively.
Substituting material values from Table~\ref{c16table} and typical values
$c\approx 0.01$ and $V\approx 5 \times 10^{-19}$ m$^{3}$, yields
$N_i\approx 10^{7}$. Thus, at this impurity level, the central limit
theorem suggests a much narrower distribution.

Nevertheless, the role of impurities in this system is not inconsequential.
While it is true that we may dispense with a time-independent distribution
of impurities in order to account for the experiments, a time-dependent change
of the impurity concentration driven by the freezing process can account for
the non-exponential behavior of the isothermal freezing data. Before we
discuss this possibility in detail, we consider the effect of impurities on
the melting temperature of the C16 droplets.

\subsection{Impurity Effects}
\label{sec:impurity}
\begin{figure}[tbp]
\epsfxsize=0.95\columnwidth
\centerline{\epsfbox{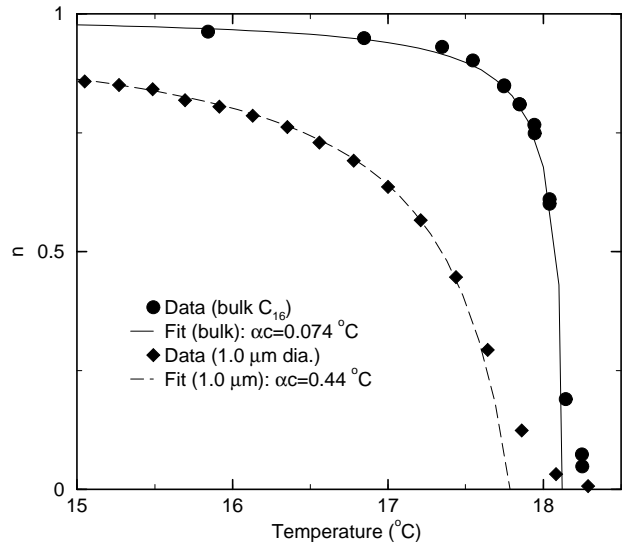}}
\caption{Melting curves obtained for bulk and emulsified $(d=1.0\ \mu m)$
C16 samples used in the nucleation study, shown as solid fraction, $n$,
versus temperature. Solid lines correspond
to best fits to Eq.~(\protect{\ref{meltcurve}}) with parameters quoted
in the legend.}
\label{melting-pic}
\end{figure}

In a liquid, the presence of impurities that are relatively insoluble in its
solid reduces its equilibrium freezing point\cite{Textbook} and broadens
its melting curve. Figure~\ref{melting-pic} shows the behavior
for both the emulsion and the bulk C16 that was used in its preparation.
Note that each emulsion droplet (or the entire bulk C16 sample) will have a
partially molten fraction, described by $x$ in the following discussion.

For molar impurity fractions $c \ll 1$, the impure liquid will be in
equilibrium with its pure solid at a suppressed temperature\cite{Textbook}
\begin{equation}
T_{m}(c)=T_{m0}-\frac{k_{{\rm B}}T}{\Delta S}c,
\end{equation}
where $T_{m0}$ is the melting point of pure C16.
A point on the melting curve at temperature $T$ reflects coexistence between
the liquid fraction $x$ with impurity concentration $c/x$ and the pure solid,
since the total number of impurities are fixed and the impurities are
assumed to be insoluble in the solid. Thus, the temperature at which the liquid
fraction is $x$ is given by
\begin{equation}
T(x)=T_{m}\left( c/x\right) =T_{m0}-\alpha c/x.
\end{equation}
Here, $\alpha \equiv k_{{\rm B}}T/\Delta S\approx 13.2$~K, obtained from the
material parameters of C16 in Table~\ref{c16table}. Inverting this
equation gives the liquid fraction $x$ as a function of $T:$%
\begin{equation}
x(c;T)=\left\{
\begin{array}{ll}
{\displaystyle {\alpha c \over T_{m0}-T}}, & T<T_{m0}-\alpha c, \\ \\
                    1, & T>T_{m0}-\alpha c.
\end{array}
\right.
\label{meltcurve}
\end{equation}
A single-parameter fit to the bulk melting curve gives excellent agreement
over the entire temperature range (see Fig.~\ref{melting-pic}),
with $c\approx 0.6\%,$ which is typical of what one might expect given
the $99\%$ purity specification of the manufacturer for the sample that was
used.

Surfactants and water have been introduced to the sample in
order to make the emulsion droplets, and therefore it is reasonable to
expect additional impurities to end up in the emulsion droplets. A similar
fit to the melting curve of emulsion droplets shown in Fig.~\ref{melting-pic}
shows an increase in the impurity concentration to about $3.3\%$. Given the high
purity and small solubility of water in C16, the surfactants are likely to be
the main source of these additional impurities.

Given that there is a range of melting temperatures, which we
attribute to varying impurity concentrations in the droplets, we propose the
following scenario in order to explain the observed behavior: The system is
presumably in thermodynamic equilibrium prior to the initiation of
nucleation events. It is reasonable to suppose that when a droplet freezes,
some of the impurities inside or at the surface of the droplet are expelled
to the
surrounding water, since the impurities are less soluble in solid than in
liquid C16. These impurities consequently redistribute themselves among
the water and remaining liquid C16 droplets in order to restore thermodynamic
equilibrium. This happens by diffusion through the continuous phase, which
occurs rapidly on the time scale over which the fraction of frozen droplets
varies. Some fraction of these impurities, determined by their relative
solubility
in water vs. C16, end up in the remaining liquid droplets, whose enhanced
impurity concentration decreases their effective undercooling and reduces
their nucleation rate. Thus, the first droplets to freeze have a {\it higher}
melting temperature than the last droplets to freeze since they have a lower
impurity concentration, which gives rise to the non-exponential behavior
during isothermal nucleation.

The theoretical analysis can easily be extended to account for such a
mechanism, which can then be compared with the experiments to test the
hypothesis.
We do this in the remainder of this section.

\subsection{Return to Isothermal Nucleation}
\label{sec:return}
\begin{figure}[tbp]
\epsfxsize=0.95\columnwidth
\centerline{\epsfbox{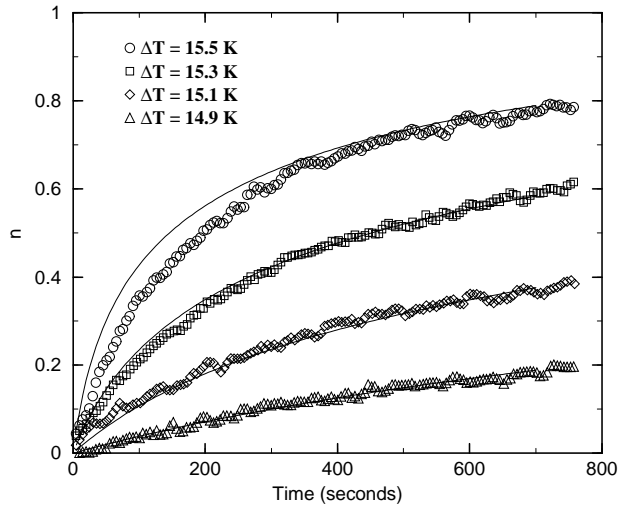}}
\caption{Solid fraction as a function of time for isothermal conditions.
Using the droplet interaction hypothesis discussed in the text we
are able to reasonably fit (solid lines) the isothermal data from a range
of undercooling values (open symbols).}
\label{isothermal-pic}
\end{figure}

The leading behavior manifested by an impurity mediated suppression of the
undercooling can be expressed as a linear decrease in the relevant melting
temperature as a function of solid C16 fraction $n$. Thus, the time evolution
of $n$ at a fixed temperature $T$ is given by
\begin{equation}
\frac{dn}{dt}=\left( 1 -n \right) AV\exp \left\{ -\frac{\Omega}{\left(
T_{m}-T-\beta n\right) ^{2}}\right\},
\label{final-diff}
\end{equation}
where $T_m$ is the {\it initial} melting temperature of the droplets and
$\beta $ is the rate at which the melting point of subsequent droplets is
reduced.
In general, $\beta$ depends on the volume fraction of the droplets in the
emulsion
and the relative solubilities of the impurities involved in the mechanism
in water and C16. Being unable to determine $\beta $ independently, we
use it as an additional fitting parameter to account for the
experimental data. A least-squares fit of the numerical integration
of Eq.(\ref{final-diff}) with respect to $\beta$, $\gamma$, and $A$ ($A$
was allowed to vary from the intial value as predicted using
Eq.(\ref{prefactor}) and the numerical constants in Table~\ref{c16table})
was performed on the data. The comparison of experiment and theory for four
different undercoolings is
shown in Fig.~\ref{isothermal-pic}. Compared to uniform nucleation or
fixed-in-time
heterogeneity assumptions, the agreement with experiment is excellent.
Furthermore,
the obtained value for $\beta $ is physically very reasonable. The best-fit
value for $\beta$ was 0.55~K, i.e., the melting point
of the last droplets to nucleate have been reduced by about half a degree.
The resulting value of the pre-exponential factor was a factor of
$10^{-1.42}$ times the classical expression.  This gave a value of
$A=2\times 10^{34}$ m$^{-3}$s$^{-1}$ at an undercooling of 15.1 K. The
resulting value for the kinetic barrier was $\gamma =9.8\pm 0.2$
mJ/m$^{2}$, in good agreement with that obtained from the linear cooling
experiments (see Sec.~\ref{sec:lincool}).

Actual values for the rates of transformation for the isothermal data in
Fig.~\ref{isothermal-pic} range from $8\times 10^{14}$ m$^{-3}$s$^{-1}$ at
$\Delta T = 14.9$~K to $7\times 10^{15}$ m$^{-3}$s$^{-1}$ at $\Delta T =
15.3$~K for early times. These rates slow down due to the impurity effect
by a factor of $\approx$ 10 after $\approx$ 60 \% conversion of the sample.

It may be noted that the freezing curve with the fastest time scale (largest
undercooling) lies somewhat below the theoretical prediction. We
attribute this discrepancy to the rapid liberation of latent heat upon the
freezing of the droplets, which raises the sample temperature and
slows the freezing of the remaining liquid droplets. Larger undercoolings
(not shown) all seem to suffer increasingly from this effect. In Fig.~\ref
{isothermal-pic} we have also adjusted the melting temperature of C16
from the value given in Table~\ref{c16table} of $T_{m0}=291.32$~K to
$T_{m}=290.95$~K.
This downward shift of the melting temperature from that of the pure
material reflects the initial impurity concentration in the emulsion
droplets prior to the freezing-induced enhancement, as well as the accuracy
in the calibration of thermistors used to measure sample temperature. (Note
that the melting temperatures for the isothermal and linear cooling
experiments were measured with different thermistors for emulsion samples
prepared at different times.)

\subsection{Return to Linear Cooling}

We may further test our hypothesis by revisiting the linear cooling problem.
If the transformation rate is slow compared to the time over
which impurity molecules are able to diffuse from one emulsion droplet to
the next,
we may assume that the impurities remain in equilibrium over the course of the
experiment.  We can estimate this equilibration time as: $ \tau_{\rm diff}
\sim %
R^2/D \left[ \left( 1 - n \right) \phi \right]^{-2/3}$ where $R$ is the
radius of
an emulsion droplet, $D$ is the diffusion coefficient for an impurity molecule
in water (which we may take to be on the order of a typical molecular
diffusivity)
and $\phi $ is the volume fraction of the emulsion droplets. We then find
that the
assumption of impurity concentration equilibration remains valid up to
transformation rates, $d n/dt$, on the order of $\tau_{\rm diff}^{-1}
\simeq 10^2 $Hz. The experimental maximum transformation rate
(taken from the maximum slope of the data in Fig.~\ref{linear-pic}) is
on the order of $10^{-4}$Hz $ \ll \tau_{\rm diff}^{-1}$.
Clearly this estimate applies only to the situation where the total
conversion to
the solid is incomplete (as is the case in our isothermal experiments --
see Fig.~\ref{isothermal-pic}) since the diffusion time to another unfrozen
droplet becomes infinite as the fraction of remaining liquid droplets ($1-n$)
drops to zero. In the linear cooling experiments where the transformation is
complete, the assumption of impurity equilibration holds until less than
$0.1$\% of the liquid droplets remain, so the observation of the breakdown
of this assumption is experimentally inaccessible.

\begin{figure}[tbp]
\epsfxsize=0.95\columnwidth
\centerline{\epsfbox{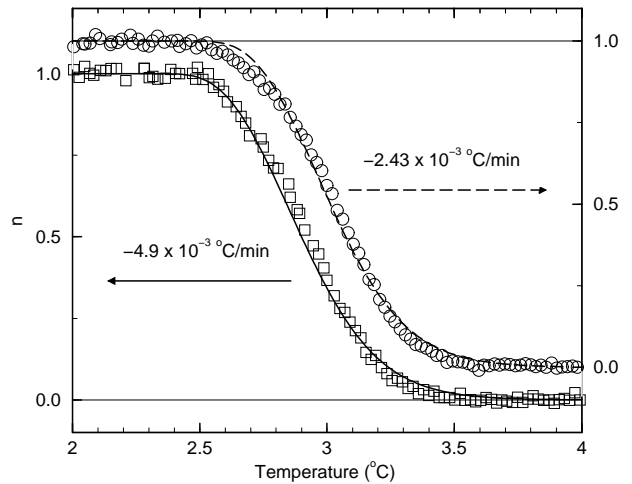}}
\caption{Solid fraction as a function of temperature at two different
linear cooling rates (open symbols) and the prediction produced by the
numerical integration of
Eq.~\protect{\ref{beta_linear_diff}} (solid lines).
}
\label{linfinal-pic}
\end{figure}

Upon modifying Eq.~(\ref{rate(t)}) to account for the suppression of the
melting temperature, time evolution of $n$ for linear cooling
is given by
\begin{equation}
\label{beta_linear_diff}
\frac{dn}{dt}= \left(1 -n \right) AVe^{-\frac{\Omega }{\left( \lambda t +
\beta n
\right)^2}}.
\end{equation}
The integration of Eq.~(\ref{beta_linear_diff}) can be performed
numerically. The
result of this integration along with the linear cooling data is shown in
Fig.~\ref{linfinal-pic} for two different cooling rates.
Once again, $\beta $ is used as a fitting parameter, with a best fit value
of $0.18$~K. Values of $A$ and $\gamma$ are as given in Sec.~\ref{sec:lincool}.

Comparing Fig.~\ref{linfinal-pic} with Fig.~\ref{linear-pic} one
immediately notes that the discrepancy in maximum slope between the
unmodified theory and the data has now disappeared. The difference in the
values of $\beta $ between the linear cooling and isothermal quench
experiments is not unexpected, given the fact that the value of $\beta $ is
sample dependent and different emulsion samples were used for the two types
of experiments. A more detailed analysis of the impurity mediated
interaction hypothesis, including a better quantitative analysis of $\beta
$ and further implications of this hypothesis on other types of experiments
is the subject of ongoing work\cite{Ertas:98}.

\section{Summary and Conclusions}
\label{sec:concl}
In this paper we report on the results of nucleation studies in C16 emulsions
during isothermal quench and linear cooling conditions. By application of
recently developed
emulsion preparation techniques to produce nearly monodisperse emulsion
samples and use of x-ray scattering for direct observation
of the crystalline phase, we greatly reduced many of the
uncertainties in the interpretation of experimental data. In doing so, we
have found potentially important effects not accounted for by the standard
interpretation of uniform nucleation. In accord with previous measurements
of this sort by
Turnbull and Cormia\cite{Turnbull:61}, the isothermal quench experiments do not
show the standard exponential growth and saturation of the fraction of solid
droplets over time.  Instead we observe a substantial decrease in the rate of
nucleation events at later times, when a significant fraction of the system has
already transformed to the solid phase. In addition, we find that in the linear
cooling experiments the maximum transformation rate predicted by the simple
nucleation theory is larger than that of the data.

Such discrepancies have previously
been attributed to the existence of some sort of temporally fixed heterogeneity
in the emulsion droplet population. One proposed source of heterogeneity
has been
the volume polydispersity in the emulsion droplets. However, using recently
developed emulsification techniques, we can better control the droplet
polydispersity and still not recover single exponential behavior. Furthermore,
by attempting to fit isothermal data at different temperatures to a
given droplet size distribution, we find that no single distribution is capable
of fitting the data at different levels of undercooling.
Along similar lines one may attempt to postulate a distribution of the
number of
heterogeneous nucleators in the droplets.  The best attempts to fit the data
with such a scheme result in unaccountably broad distributions of impurity
concentrations in the emulsion droplets. Such distributions are highly
unlikely in
light of the central limit theorem.

In contradistinction to both of these ideas, we propose that the observed
heterogeneity in the emulsion droplets is dynamically generated by the
nucleation
process itself. Freezing droplets expel impurities (some of which might be
attributed to the addition of surfactant) which then go into solution in the
remaining liquid C16 droplets in order to restore thermodynamic equilibrium.
By doing so, these mobile impurities reduce the melting temperature of the
remaining liquid droplets, which experience a reduced thermodynamic driving
force towards solidification and consequently a reduced nucleation rate.
This proposal is able to account for both the isothermal and linear cooling
data at
a variety of undercoolings and cooling rates, respectively. To a good
approximation,
the magnitude of this effect can be parametrized by a sample dependent
parameter $\beta $, which expresses the rate at which  the melting point
of remaining liquid droplets is suppressed as a function of the solid
droplet fraction.  To obtain
accurate quantitive measurements of either the activation barrier toward
nucleation
or the pre-exponential factor, this effect needs to be taken into account.
Note that measurements described in this paper were performed on emulsion
samples prepared with a mixture of anionic surfactants. Measurements of a
C16 emulsion sample prepared with a single nonionic surfactant does not
appear to show such a large impurity effect \cite{Ertas:98}.

Further tests of this theory can be anticipated. It should be possible to
directly detect and measure the suppression of melting temperature by an
experiment where the emulsion is partially melted and subsequently refrozen
at a temperature where no new nucleation is allowed. Such careful studies
of the remelting process coupled to a more detailed theory for the impurity
expulsion and transport should lead to a better understanding of whether
and how the dynamics of the freezing process effects the results of
emulsion nucleation studies. Additional work along these lines is currently
in progress\cite{Ertas:98}.

We would like to thank H.~Shao and H.~Gang for their help in the initial
stages of the
experiments, J.~Wang for performing the isothermal quench experiments, W.
Gordon for his assistance with the experiments, and
E.~Sirota, S.~Milner and J.~Hutter for useful discussions.  For work done
while at the University of Pennsylvania, A.~L.
acknowledges support by  the Donors of The Petroleum Research Fund 
administered by the 
American Chemical Society.


\begin{references}
\bibitem[*]{home}
Present address: Department of Physics \& Astronomy, University of Pennsylvania,
Philadelphia, PA 19104.

\bibitem{Vonnegut:48}B. Vonnegut, J. Colloid Sci. {\bf 3}, 563 (1948).

\bibitem{Turnbull:52}D. Turnbull, J. Chem. Phys. {\bf 20}, 411 (1952).

\bibitem{Turnbull:61}D. Turnbull and R. L. Cormia, J. Chem. Phys.
{\bf 34}, 820 (1961).

\bibitem{Kelton:91}K.~F.~Kelton, Solid State Physics {\bf 45}, 75 (1991).

\bibitem{Perepezko:97}J.~H.~Perepezko, Materials Science and Engineering
{\bf A226-228}, 374 (1997).

\bibitem{Perepezko:84}J.~H.~Perepezko, Materials Science and Engineering
{\bf 65}, 125 (1984).

\bibitem{Bibette:91}J. Bibette, J. Colloid Interface Sci. {\bf 147}, 474 (1991).

\bibitem{Kandori91a}K. Kandori, K. Kishi and T. Ishikawa, Colloids
and Surfaces {\bf 55}, 73 (1991).

\bibitem{Kandori:91}K. Kandori, K. Kishi and T. Ishikawa, Colloids
and Surfaces {\bf 61}, 269 (1991).

\bibitem{Turnbull:78}D. Turnbull and F. Spaepen, J. Polym. Sci.,
Polym. Symp. {\bf 63}, 237 (1978).

\bibitem{Uhlmann:75}D. R. Uhlmann, G. Kritchevsky, R. Straff
and G. Scherer, J. Chem. Phys. {\bf 62}, 4896 (1975).

\bibitem{Oliver:75}M.~J.~Oliver and P.~D.~Calvert, J. Crystal
Growth {\bf 30}, 343 (1975).

\bibitem{McClements:90}D.~J.~McClements; E.~Dickinson and
M.~J.~W.~Povey, Chem. Phys. Lett. {\bf 172}, 449 (1990).

\bibitem{Dickinson:93}E.~Dickinson, F.-J.~Kruizenga,
M.~J.~W.~Povey and M.~v.~d.~Molen, Col. and Surf. A {\bf 81}, 273 (1993).

\bibitem{Dickinson:90}E.~Dickinson, M.~I.~Goller, D.~J.~McClements,
S.~Peasgood and M.~J.~W.~Povey, J. Chem. Soc. Faraday Trans. {\bf 86}, 1147
(1990).

\bibitem{Dickinson:91} E.~Dickinson, D.~J.~McClements and M.~J.~W.~Povey,
J. Colloid Interface Sci. {\bf 142}, 103 (1991).

\bibitem{Bibette:1991}J.~Bibette, Journal of Colloid and Interface
Science {\bf 147}, 474 (1991).

\bibitem{tableref} D.~M.~Small, {\it The Physical Chemistry of Lipids:
>From Alkanes to Phospholipids}, Plenum Press (New York, 1986).

\bibitem{Krafftpoint} H.~Nakayama, K.~Shinoda and E.~Hutchinson, J. Phys.
Chem. {\bf 70}, 3502 (1966).

\bibitem{EOref} M.~Hato and K.~Shinoda, J. Phys. Chem. {\bf 77}, 378 (1973).

\bibitem{HenryShao} H.~Shao, Ph.D Thesis, {\it X-ray Scattering Study of
Structures and Phase Transitions of Normal Alkanes}, The Ohio State
University (1995).

\bibitem{Turnbull:49}  D.~Turnbull and J.~C.~Fisher, J. Chem. Phys.
{\bf 17}, 71 (1949).

\bibitem{cp-note}  We neglect the difference in specific heats
$C_{{\rm P}}$ between the liquid and solid phases in this case.

\bibitem{Wu:96}  D.~Wu, Solid State Physics {\bf 50}, 37 (1996).

\bibitem{Abramowitz:72}  M.~Abramowitz and I.~A.~Stegun,  {\it Handbook of
Mathematical Functions\/}, p. 260, Dover Publications, Inc. (New York, 1972).


\bibitem{Turnbull:69} D.~Turnbull, Contemp. Phys. {\bf 10}, 473 (1969).

\bibitem{growth-note}At typical crystal growth velocities of meters per second,
the time lag between nucleation and completion of growth in a single emulsion
droplet is on the order of a microsecond. This estimate is not
substantially changed
by the fact that these materials form plate-like structures during freezing.
Even though the formation of plates suggests slower growth along certain
cystallographic directions, the ratio of the slow rate to the fast one would
have to be less than $10^{-6}$ to have an effect on the experiments.

\bibitem{Mason:98} D.~B.~Siano, J. Chem. Educ. {\bf 49}, 755 (1972).

\bibitem{emulsionsize-note}  Based on the estimate that the standard
deviation for the droplet diameter is about $0.1$, the standard deviation
for the emulsion droplet volume distribution should be about $0.3$.

\bibitem{Textbook} E.~M.~Lifshitz and L.~P.~Pitaevskii, {\it Statistical
Physics, 3rd Edition, Part 1\/}, Chap. 9, Pergamon Press (New York, 1980).

\bibitem{Ertas:98} D.\ Erta{\c s} {\it et. al.\/}, in preparation.

\end{references}
\end{document}